# Field test of quantum key distribution in the Tokyo QKD Network

M. Sasaki[1,*], M. Fujiwara[1], H. Ishizuka[1], W. Klaus[1], K. Wakui[1], M. Takeoka[1],
A. Tanaka[2], K. Yoshino[3], Y. Nambu[3], S. Takahashi[2], A. Tajima[2], A. Tomita[4],
T. Domeki[5], T. Hasegawa[6], Y. Sakai[6], H. Kobayashi[6], T. Asai[6], K. Shimizu[6],
T. Tokura[6], T. Tsurumaru[6], M. Matsui[6], T. Honjo[7], K. Tamaki[7], H. Takesue[7],
Y. Tokura[7], J. F. Dynes[8], A. R. Dixon[8,9], A. W. Sharpe[8], Z. L. Yuan[8], A. J. Shields[8],
S. Uchikoga[8], M. Legré[10], S. Robyr[10], P. Trinkler[10], L. Monat[10] and J.-B. Page[10],
G. Ribordy[10], A. Poppe[11], A. Allacher[12], O. Maurhart[11], T. Länger[11], M. Peev[11],
and A. Zeilinger[12,13]

[1]*National Institute of Information and Communications Technology, 4-2-1 Nukui-kitamachi, Koganei, Tokyo 184-8795, Japan*

[2] *System Platforms Research Laboratories, NEC Corporation, 1753 Shimonumabe, Nakahara, Kawasaki, Kanagawa 211-8666, Japan*

[3] *Green Innovation Research Laboratories, NEC Corporation, 34 Miyukigaoka, Tsukuba, Ibaraki 305-8501, Japan*

[4] *Graduate School of Information Science and Technology, Hokkaido University, Kita 14 Nishi 9, Kita-ku, Sapporo, Hokkaido 060-0814, Japan*

[5] *Network Platform Business Division, NEC Communication Systems, 3-4-7 Chuo, Aoba, Sendai, Miyagi 980-0021, Japan*

[6] *Information Technology R&D Center, Mitsubishi Electric Corporation, 5-1-1 Ofuna, Kamakura, Kanagawa 247-8501, Japan*

[7] *NTT Basic Research Laboratories, NTT Corporation, 3-1 Morinosato, Wakamiya, Atsugi, Kanagawa 243-0198, Japan*

[8] *Toshiba Research Europe Ltd, 208 Cambridge Science Park, Cambridge CB4 0GZ, United Kingdom*

[9] *Cavendish Laboratory, University of Cambridge, J. J. Thomson Avenue, Cambridge CB3 0HE, United Kingdom*

[10] *ID Quantique SA, Chemin de la Marberie 3, 1227 Carouge/Geneva, Switzerland*

[11] *Austrian Institute of Technology GmbH, Donau-City-Straße 1, 1220 Vienna, Austria*

[12] *Faculty of Physics, University of Vienna, Boltzmanngasse 5, 1090 Vienna, Austria*

[13] *Institute for Quantum Optics and Quantum Information, Austrian Academy of Sciences, Boltzmanngasse 3, 1090 Vienna, Austria*

[*]*psasaki@nict.co.jp*

**Abstract:** A novel secure communication network with quantum key distribution in a metropolitan area is reported. Different QKD schemes are integrated to demonstrate secure TV conferencing over a distance of 45km, stable long-term operation, and application to secure mobile phones.

©2010 Optical Society of America

**OCIS codes:** 270.0270 Quantum optics; 270.5568 Quantum cryptography.

# 1. Introduction

Among all the methods of encryption ever devised, only one has been proven to be information-theoretically secure, i.e. secure against an eavesdropper who has unbounded ability. It is the one-time pad (OTP). The key should be used only once and be as long as the message to be sent. The efficient distribution of such long keys remains an issue. Quantum key distribution (QKD) provides a means to deliver key material for OTP over an optical network. Experimental demonstrations and development of QKD were carried out by many research institutes in the 1990s. In the 2000s QKD systems were transferred from the controlled environment of laboratories into a real-world environment for practical use [1]. Progress has also been made in theory, not only developing new tools to prove protocols themselves but also analyzing the security of practical QKD systems [2]. The commercialization of QKD has also been successful [3-5].

In the past decade, multi-user QKD networks have been extensively investigated in field environments. The DARPA Quantum Network, as part of a project supported by the US Defense Advanced Research Projects Agency (DARPA), pioneered the deployment of QKD in a field network [6]. The network consists of 10 nodes linked together through an actively switched optical network. The European FP6 project Secure Communication using Quantum Cryptography (SECOQC) integrated a number of different QKD systems into one quantum backbone (QBB) network, developing a cross-platform interface [7]. From the SECOQC project, the European Telecommunications Standards Institute (ETSI) industry specification group for QKD (ISG-QKD) was launched to offer a forum for creating universally accepted QKD standards [8]. Long-term QKD operation also has been tested in a field environment, such as the Swiss Quantum network in Geneva [9] the Durban network in South Africa developed by the Durban–Quantum City project [10] and the Cambridge Network [11]. Transparent network implementations of QKD have been demonstrated, such as a dynamically reconfigurable network in a testbed of the Advanced Technology Demonstration Network (ATDNet) in the Washington D.C. area by Telcordia Technologies [12], a passive optical network consisting of core ring and access network in Madrid by Universidad Politécnica de Madrid and Telefónica Investigación y Desarrollo [13], a hierarchical network consisting of a 5-node wavelength division multiplexing (WDM) quantum backbone and subnets connected by trusted nodes, in Wuhu, Anhui, by one group from the University of Science and Technology of China [14], and an all-pass optical switching network in Hefei, Anhui, by another group from the same university [15].

All these QKD field trials adopted either or both of the following two kinds of networking schemes: key relay via trusted nodes, and transparent link via optical switching. The former requires guaranteed physical security of the relay nodes, but can expand key distribution distance arbitrarily. The latter can realize key establishment for many users with less complexity of key management over an untrusted network. The distance and the secure key rate are, however, limited by an overall optical loss. Which networking scheme is mainly used depends on the purpose and the infrastructure to be installed.

To appeal to a wider adoption of QKD, it is indispensable to demonstrate applications demanded by potential users of high-end security technology, such as secure TV conferencing, and secure mobile phone in an area as wide as possible. The typical QKD link performance in the field networks so far was represented by the secure key rate of a few kbps at a distance of a few tens of km, which was only sufficient to encrypt voice data by real time OTP, or to feed the primary session key to a classical encryptor. The secure key rate needs to be significantly improved for other applications. To expand the QKD distance, currently one needs to rely on a key relay via trusted nodes. In this paper, we present a metropolitan QKD network with trusted nodes, called the Tokyo QKD Network, where the latest QKD technologies and an upgraded application interface were installed, and various applications including secure TV conferencing and secure mobile telephony were demonstrated.

## 2. Outline of the Tokyo QKD Network

The Tokyo QKD Network consists of parts of the NICT open testbed network called Japan Giga Bit Network 2 plus (JGN2plus), as shown in Fig. 1(a). It has four access points, Koganei, Otemachi, Hakusan, and Hongo. The access points are connected by a bundle of commercial fibers. They include many splicing points and connectors, even run partly through the air over utility poles. The percentage of aerial fibers is about 50% causing the links to be quite lossy and susceptible to environmental fluctuation. The loss rate is roughly 0.3dB/km on average for the Koganei-Otemachi link, and as high as 0.5 dB/km on average for the other two links. The fibers are also noisy, i.e. photon leakage from neighboring fibers causing inter-fiber crosstalk in the same cable is often observed [16].

(a)

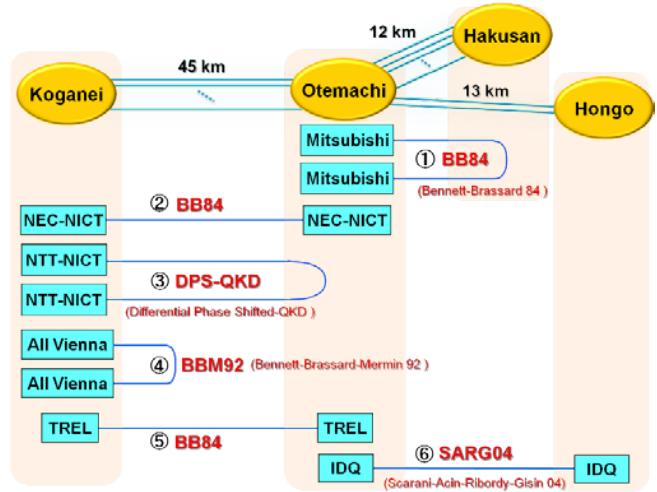

(b)

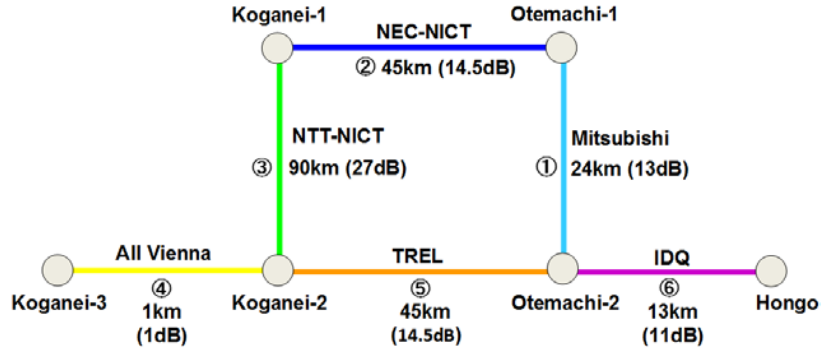

Fig. 1. (a) Physical link configuration of the Tokyo QKD Network. It is a mesh-type network consisting of four access points, Koganei, Otemachi, Hakusan, and Hongo. In total, six kinds of QKD systems are installed in these access points. Some QKD links are connected in a loop-back configuration with parallel fibers. (b) Logical link configuration with 6 nodes. Link distances range from 1 km to 90 km.

In 2010, nine organizations from Japan and the EU participated in the Tokyo QKD Network operation. On the Japanese side these were, besides NICT, the three companies NEC, Mitsubishi Electric Corporation (Mitsubishi) and NTT, and on the EU side, Toshiba Research Europe Ltd. (TREL) from the UK, ID Quantique (IDQ) from Switzerland and three organizations from Austria, the Austrian Institute of Technology (AIT), the Institute of Quantum Optics and Quantum Information (IQOQI) and the University of Vienna, forming together one team dubbed "All Vienna". Using their QKD devices, a total of six QKD links were formed as shown in Fig. 1(a). Link no. 1 by Mitsubishi used decoy-state BB84 protocol [17] in a 24 km link in the loop back configuration between Otemachi and Hakusan. Link no. 2 by NEC used decoy-state BB84 protocol in a 45 km link between Koganei and Otemachi, with NICT's superconducting single photon detector (SSPD). Link no. 3 by NTT used DPS-QKD [18] and took on the challenge of long distance QKD over 90 km in a loop back configuration also using NICT's SSPD. Link no. 4 by All Vienna demonstrated BBM92 [19] with installed fibers in the premises of NICT. Link no. 5 by TREL demonstrated their decoy-state BB84 system using electrically cooled self-differencing avalanche photodiodes (SD-APDs), in a 45 km link. Finally, link no. 6 by IDQ applies their commercial system using the SARG04 protocol [20] in the Otemachi-Hongo link. This link configuration logically forms a 6-node mesh-type network, as shown in Fig. 1(b).

We adopted a three-layer architecture based on the key relay via trusted nodes, as shown in Fig. 2, similar to the SECOQC network [7]. The quantum layer consists of point-to-point quantum links, forming the QBB. Each link generates the secure key in its own way. The QKD protocols as well as the format and size of the key material can be arbitrary. QKD devices push the key materials to the middle layer, called the key management layer. In this layer, a key management agent (KMA) located at each site receives the key material via the application interface (API). Each KMA is actually a PC, which is physically protected, and works as a trusted node. Networking functions are performed entirely in this KM layer by software. A KMA can relay a secure key shared with one node to a second node by OTP-encrypting the key, using another key shared with the second node. Thus a secure key can be shared between nodes that are not directly connected to each other by a quantum link. In the communication layer, secure communication is ensured by using the distributed keys for encryption and decryption of text, audio or video data produced by various applications. Finally, a key management server (KMS) is introduced for a centralized management of the key life cycle and the provision of secure paths. This differentiates the Tokyo QKD Network from the SECOQC network. In the latter, the secure path-finding problem is solved by an autonomous search algorithm following standard approaches in networking. The main reason for adopting the centralized management in the Tokyo QKD Network is that it assumes a test case of a government-chartered network or a mission critical infrastructure network which often have a central dispatcher or a central data sever.

An important task of the Tokyo QKD Network trial was to establish a common API which provides the interoperability of a great variety of different QKD devices. NEC and NICT developed an API which is compatible with the SECOQC QBB Link Interface (QBB-LI), i.e. the original Japanese API software uses function calls that are similar to those used by the open source SECOQC QBB-LI. This software compatibility allowed for a smooth interconnection between the Japanese and European QKD systems. The API was installed into the KMAs. Each KMA has the job to resize and save the key material as well as to store link statistic relevant data such as quantum bit error rate (QBER) and key generation rate. It furthermore forwards all these data to the KMS who coordinates and oversees all links in the network.

The physical implementation of the three-layer architecture is depicted by the wiring diagram in Fig. 3. The blue lines represent optical fibers for the quantum layer. For links no. 1, 3, 5 and 6 a second fiber was used to send the classical information required for the QKD protocol and synchronization signals, according to the original specification of the Tokyo QKD network. For links 2 and 4, the classical information was sent through a second fiber in the local area network (shown as a different color in Fig.3.) All the KMAs are located inside the network isolated from each other by L2 switches. The network itself is connected to the Internet via a router.

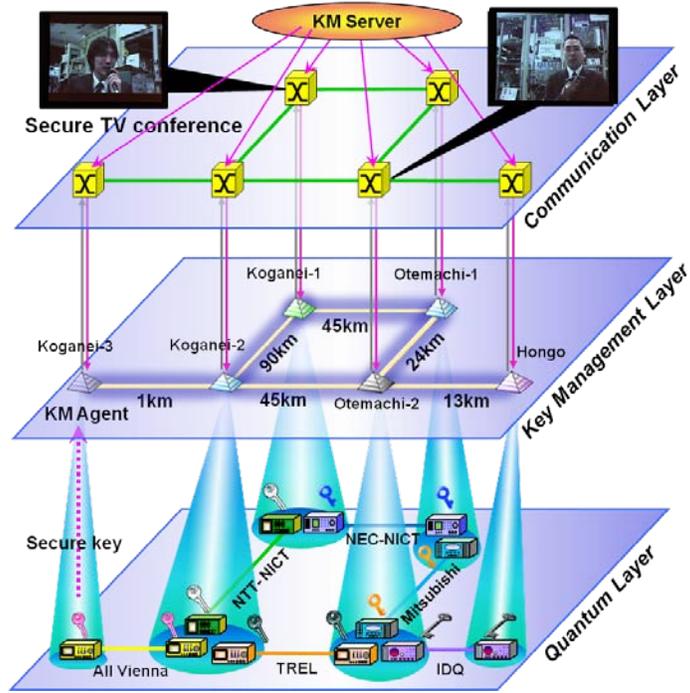

Fig. 2. Three-layer architecture of the Tokyo QKD Network. It consists of the quantum, the key management, and the communication layer.

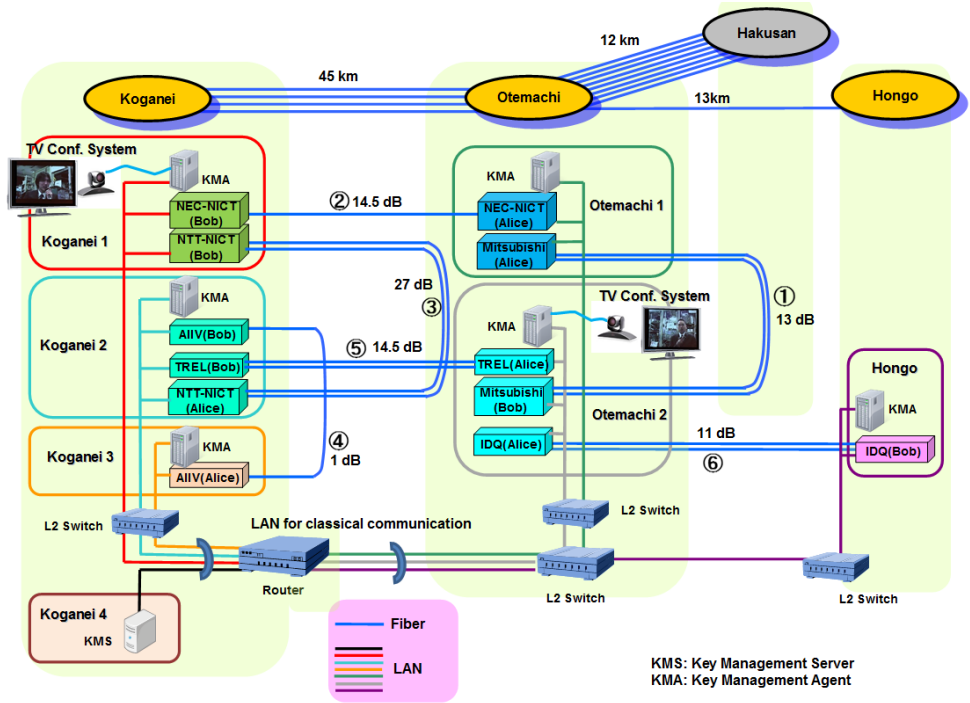

Fig. 3. Wiring diagram of the Tokyo QKD Network.

## 3. QKD systems used in the Tokyo QKD Network

The performance of QKD has been much improved in recent years owing to the progress in state-of-art technologies, such as novel photon detectors operating at higher speed with lower noise and faster electronics. In the Tokyo QKD Network, high speed QKD systems developed by NEC-NICT and TREL have enabled real-time secure video conferencing in a metropolitan area. The DPS-QKD system developed by NTT allows for real-time long-distance secure voice communication. Mitsubishi combined its QKD system with an application for secure telephony over smartphones. A reliable and highly stable commercial system was demonstrated by IDQ. Finally, All Vienna contributed with an impressive next-generation QKD system using quantum entanglement.

*3.1 NEC-NICT system*

NEC and NICT developed a one-way decoy-state BB84 system, aiming at fast QKD for metropolitan-scale distances, which can realize OTP encryption of video data. The system is designed for a multi-channel QKD scheme with wavelength division multiplexing (WDM). Each channel is operated at a clock rate of 1.25 GHz. A block diagram of the QKD system is depicted in Fig. 4. The upper part shows the optical transmission block and the lower part represents the key distillation block, where a dedicated hardware engine is used for the key distillation process. The hardware engine has a large memory, large-size field programmable gate arrays (FPGAs), and high speed in/out interfaces, which can potentially handle up to 8 WDM channels, i.e. for a processing speed of up to 10Gbps.

Figure 5 shows the photon transmission setup of the NEC-NICT QKD system. In the transmitter, a laser diode produces 1550 nm photon pulses with 100 ps width at a repetition rate of 1.25 GHz. A 2-by-2 asymmetric Mach-Zehnder interferometer (AMZI) made of a polarization free planar-lightwave-circuit (PLC) splits these pulses into pairs of double pulses with a 400 ps delay. A dual-drive Mach-Zehnder modulator produces four quantum states in the time-bin encoding, according to pseudorandom numbers provided by a controller. The quantum signal is combined with the clock and frame synchronization signals by a WDM coupler, and these multiplexed signals are transmitted through the same fiber for precise and automatic synchronization [21].

In the receiver, the quantum and the synchronization signals are divided by a WDM filter. The quantum signal is discriminated by a 2-by-4 asymmetric and totally passive PLC-MZI, and is then detected by a four-channel SSPD, which is free from the afterpulse effect and complex gate timing control. The detection efficiency and the dark count rate of the SSPD itself are about 15% and 100 cps, respectively [22]. When it is combined with the QKD system, however, the total detection efficiency reduces to about 7% because the active time window imposed on the time-bin signal cannot cover the whole pulse spreading after the fiber transmission, and the noise count rate increases to 500 cps due to stray light.

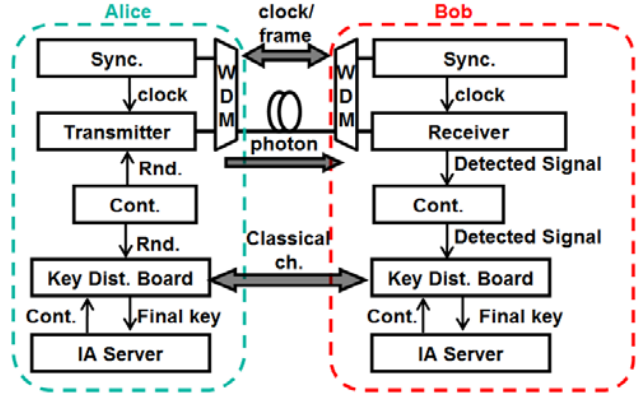

Fig. 4. Block diagram of NEC's QKD system. Sync.: Synchronization, Cont.: Controller, Rnd.: Random Number, Dist.: Distillation, IA: Intel Architecture.

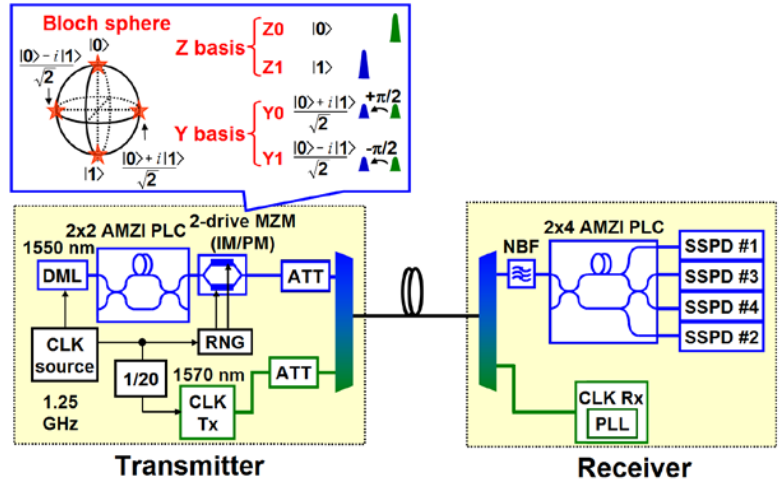

Fig. 5. Photon transmission schematics of the NEC-NICT system. AMZI: Asymmetrical Mach-Zehnder Interferometer, PLC: Planar Lightwave Circuit, DML: Directly Modulated Laser, IM/PM: Intensity Modulation/Phase Modulation, CLK: Clock, RNG: Random Number Generator, MZM: Mach-Zehnder Modulator, IM: Intensity Modulation, PM: Phase Modulation, ATT: Attenuator, NBF: Narrow Bandpass Filter, PLL: Phase Locked Loop, SSPD: Superconducting Single Photon Detector.

The key distillation engine performs frame synchronization, sifting, random permutation (RP), error correction (EC) and privacy amplification (PA). The sifted key is processed in block units of 1 Mbit, and RP, EC and PA are executed in real time, within 200 ms, at this block size. Low density parity check code with 1 Mbit code length is implemented for EC. The coding rate can be adjusted at an appropriate value depending on the quantum bit error rate (QBER), such as 0.75, 0.65, or 0.55 for QBER <3.5%, 5.5%, or 7.5%. This means that PA after EC is performed with a modified Toeplitz matrix [23] of a block size of 750 kbit, 650 kbit, or 550 kbit.

Figure 6 shows the temporal fluctuation of the measured QBER and the sifted and final key rates after 45 km transmission (14.5 dB channel loss), where a single channel out of 8 channels was working with the SSPD. The averaged photon numbers of signal and decoy pulse were 0.5 and 0.2 photons/pulse, respectively. The averaged QBER was 2.7% and the averaged sifted key rate was 268.9 kbps. The processing time of EC and PA at this error rate were confirmed as 170 ms and 95 ms, respectively. NEC estimated the leaked information by applying the decoy state method analysis [24, 25] with an asymptotical estimation, resulting in the averaged final secure key rate of 81.7 kbps.

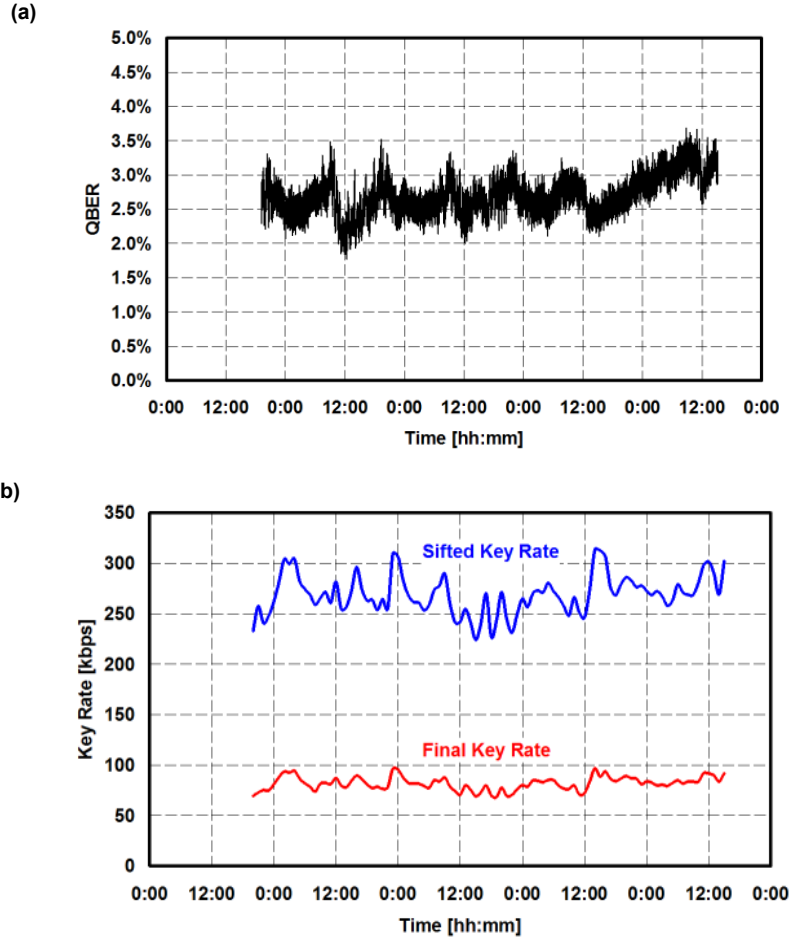

Fig. 6 (a) Measured quantum bit error rate, and (b) sifted and secure (final) key rates.

*3.2 TREL system*

The TREL system is based on a one-way, GHz clocked decoy-state BB84 scheme [26]. Fig. 7 outlines the system. Inside the transmitter (Alice), a distributed feedback laser (DFB) pulsed at 1 GHz produces 1550 nm photon pulses with a 50 ps width. An intensity modulator creates three different pulse intensities for the decoy scheme; signal pulses of 0.5 photons/pulse sent with almost 99% probability, and two different decoy pulses of 0.1 and 0.0007 photons/pulse sent with less than 1% probabilities. The information is encoded on the photon's phase using an AMZI, with a phase modulator located in one arm.

In the receiver (Bob), the photons pass through an electronic polarization controller (EPC), to correct for any polarization drift in the fiber, and then Bob's interferometer. One arm contains a phase modulator for phase decoding while the other arm contains a fiber stretcher which compensates drift in the fiber length. The photons are detected by electrically cooled (-30°C) InGaAs APDs in self-differencing (SD) mode [27, 28] at 1 GHz, with a detection efficiency of ~19% and a dark count rate of approximately 10 kHz. Afterpulse noise is the dominant contribution to the QBER which averages around 3.8%, but is strongly suppressed by the self-differencing technique.

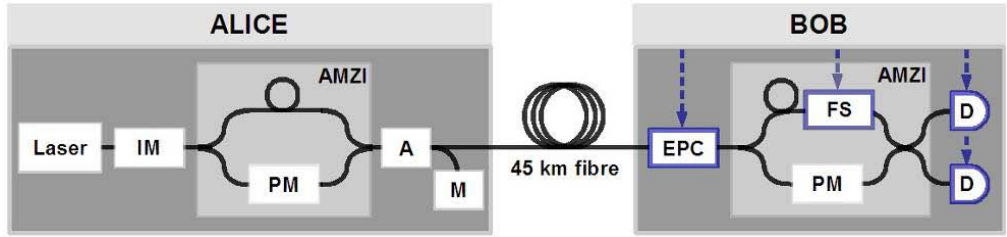

Fig. 7. Schematic of the Toshiba GHz QKD system. IM=intensity modulator; PM=phase modulator; A: attenuator, M: intensity monitor, EPC: electronic polarization controller, FS: fiber stretcher, D: SD-APD detector, AMZI: asymmetric Mach-Zehnder interferometer.

The detector count rates are used as a feedback signal to adjust the delay position of the detector gate as well as the polarization controller state [29]. The fiber stretcher is controlled by using the QBER as a feedback signal. This kind of active stabilization scheme is crucial to minimize the effect of secure key rate reduction due to the finite key size.

The sending and receiving QKD units are kept in synchronicity by a 1550nm clock laser that is multiplexed on the classical communication channel. Both this data (classical) fiber and quantum fiber run together in the same fiber bundle. This is advantageous as fluctuations in the ambient weather conditions results in temperature drifts and vibrations affecting both fibers in a similar manner. Consequently, the relative temporal drift between Alice and Bob is quite slow, i.e. tens of minutes.

We note that gated single photon detectors are particularly advantageous for QKD over noisy fibers. The detectors' active temporal width is only 100 ps and coupled with gated operation most of the stray photons in the fiber are rejected. Hence it was not necessary to use a band pass filter at Bob. Additionally a fiber Bragg grating at Alice ensured the temporal width of the quantum signals was recovered at Bob's detectors.

Sifting, EC and PA communications are wavelength multiplexed with the clock fiber, thus avoiding latency problems associated with the local area network. EC and PA are performed using multi-threaded applications on two multi-core processors located at Alice and Bob. Each EC thread works on a 1Mbit bit sequence using the cascade algorithm [30], while the block size for PA was several hundred kilobits using a Toeplitz matrix approach. Tests using 4-core processors reveal EC speeds in excess of 5.5 Mbps. These speeds are adequate for GHz QKD systems with channel loss exceeding 10dB.

The above novel technologies account for the dramatic increase in secure bit rates of decoy-state BB84, as shown in Fig. 8. Figure 8(a) indicates a secure bit rate of >1 Mbps obtained over a 50km spool of single mode fiber (loss 10dB) in the laboratory [26]. Figure 8 (b) shows the variation in secure bit rate obtained for the Otemachi-Koganei 45km link in the field test, performed in early October 2010. An average secure bit rate of 304 kbps over a 24 hour segment was obtained despite the high channel loss of 14.5dB. The system operated continuously until after the network launch date on 18$^{th}$ October 2010. The bit rates obtained in both the laboratory and field test trials allow secure OTP video transmission.

Remarkably, despite operating over fibers with higher loss, we have been able to increase the bit rate by over two orders of magnitude compared to results we obtained in the previous SECOQC QKD field test [7]. This is by far the highest sustained QKD bit rate produced in an installed fiber network to date.

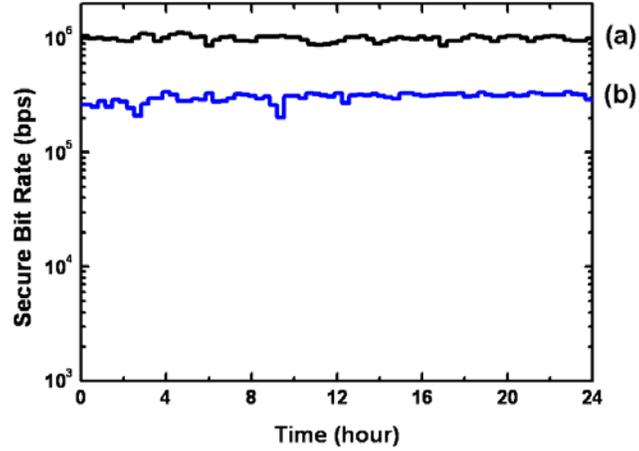

Fig. 8. (a) Secure bit rate for a 50 km spool of single mode fiber recorded over a 24 hour duration (from [26]). (b) Secure bit rate for the 45 km fiber link between Otemachi and Koganei recorded over a 24 hour duration.

*3.3 NTT-NICT system*

NTT and NICT developed a 1GHz-clocked differential phase shift QKD (DPS-QKD) system for long-distance QKD. The link has a total length of 90 km going from Koganei to Otemachi and back. The DPS-QKD scheme is especially suitable for fiber transmission [18, 31], and is known to be secure against general individual attacks [32]. Figure 9 shows the setup. In the transmitter, a 1551 nm continuous light wave from a semiconductor laser is changed into a 70 ps width pulse stream with a 1 GHz repetition rate using a $LiNbO_3$ intensity modulator. Each pulse is randomly phase-modulated by $\{0, \pi\}$ with a $LiNbO_3$ phase modulator driven by the random bit signal from the FPGA board [33]. The optical pulse is attenuated to 0.2 photons/pulse and then transmitted to Bob through an optical fiber acting as quantum channel with a total loss of 27dB. The 100 kHz synchronization signal, which makes up the head of a 10 kbit block of random bits, is also generated by the FPGA board, and converted into a 1560 nm optical pulse by a distributed feedback laser with an electro-absorption modulator (EA-DFB), and is then sent over another optical fiber whose total loss is 30.0dB.

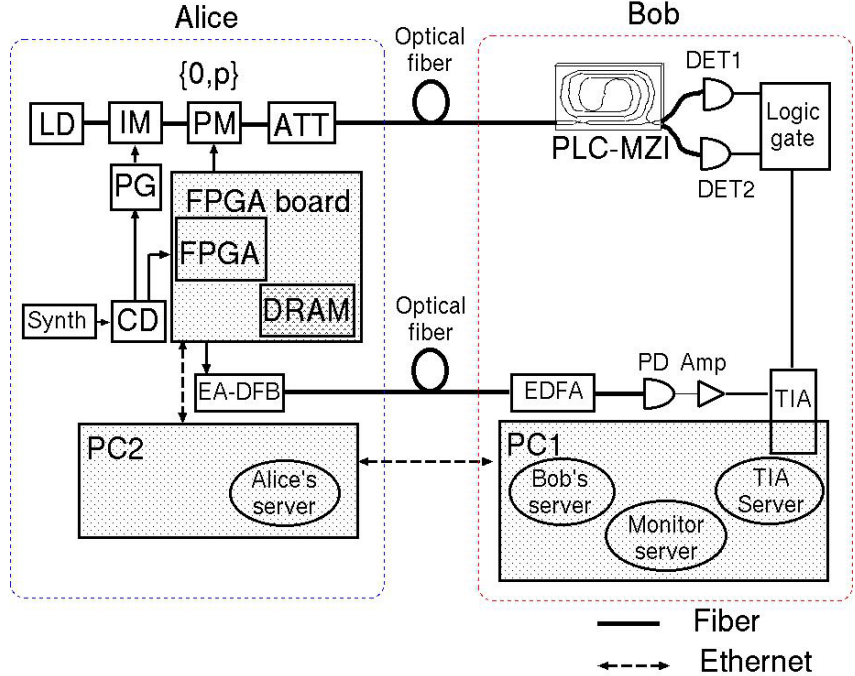

Fig. 9. Experimental setup of DPS-QKD system. LD: laser diode, IM: intensity modulator, PG: pulse generator, PM: phase modulator, ATT: attenuator; PLC-MZI: planar lightwave circuit Mach-Zehnder interferometer, Synth: synthesizer; CD: clock divider, FPGA: field programmable gate arrays, PD: photodiode, TIA: time interval analyzer, PC: personal computer.

In the receiver, the 1 GHz pulse stream is input to a PLC-MZI. The output ports of MZI are connected to SSPDs whose detection efficiencies and dark count rates are about 15% and about 100 cps, respectively. The detected signals are input into a time interval analyzer (TIA) via a logic gate to record the photon detection events. The optical synchronization pulses are first amplified by an EDFA, then received by a photo detector (PD), and finally are used as a reference time in the TIA.

The detection events from the TIA are sent to Bob's server via the TIA server. Bob's server generates his sifted key and sends the time information to Alice's server, through an Ethernet connection. Alice's server generates her key with the phase modulation information obtained from the DRAM on the FPGA board through the gigabit Ethernet interface, and the time information from Bob's server. Both servers send 10% of their keys to a monitor server, which estimates the key generation rate and QBERs. The remaining 90% are sent to the key distillation engine developed by NEC, which performs error correction and privacy amplification to distill the secure keys in the same way as in the NEC-NICT system.

First, we checked the stability of the sifted key generation. Figure 10(a) shows the experimental results. Ultra stable sifted key generation for more than 8 days was demonstrated. The spike-like degradation of QBER was caused by the eavesdropping demonstration during the UQCC2010 conference [34]. Sifted key generation rate and QBER were about 18 kbps and 2.2% on average, respectively. Next, we performed secure key generation experiment combining with the key distillation engine. Figure 10(b) shows the experimental results. A stable operation for about 4 hours was demonstrated. Secure key generation rate was about 2.1 kbps on average. The distilled secure keys are secure against general individual attack. The sifted key generation rate and QBER were about 15 kbps and 2.3% on average, respectively. This allows for OTP encryption of voice data in real time, even over a distance of 90 km.

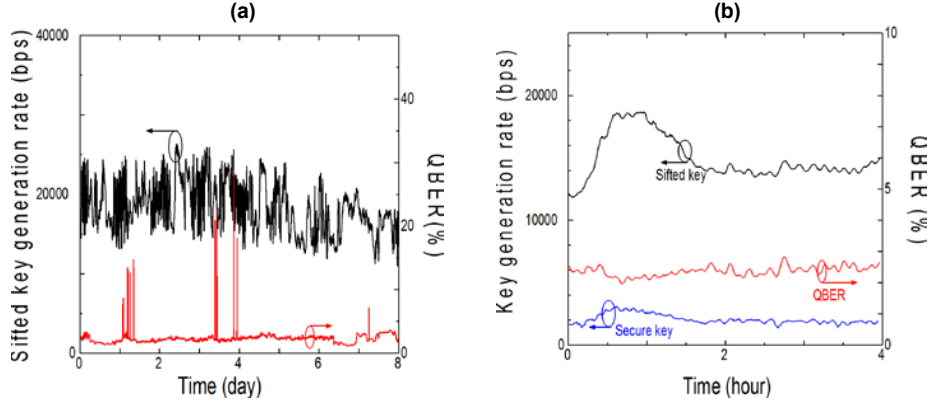

Fig. 10. Experimental results of (a) sifted key generation and (b) secure key generation rates.

*3.4 Mitsubishi system*

Mitsubishi applied a decoy-state BB84 system to the 24 km link with a total loss of 13dB. A schematic diagram of the setup is shown in Fig. 11. The experimental setup is one-way scheme based on 500 MHz Mach-Zehnder interferometers (MZIs). In this system, a laser source with a pulse width of 500 ps is operated at 100MHz. The quantum and classical light sources are designed using DWDM (dense wavelength division multiplexing) DFB laser modules at telecom wavelengths (quantum signal: 1549.32 nm; classical signal 1550.92 nm).

The use of PLCs and polarization controllers results in a highly stable system. Time synchronization was done by classical signals through a second optical fiber. In this field test, the polarization property of the channel was relatively stable. In the transmission line with the largest polarization fluctuations such as aerial fibers, the QKD bit rate was stabilized by monitoring the high-intensity classical light with our WDM/DEMUX modules and by compensating the polarization of the quantum signal with our polarization compensation module. The DWDM DEMUX module was also designed for providing channel isolation between the two signals of more than 80dB.

The system uses light pulses with four different intensity levels (signal: 0.7 photons per pulse; decoy: 0.3, 0.1 and vacuum). It consists of PLCs with polarization stabilizers and commercial APDs with a detection efficiency of 3% and a dark count probability of $6\times10^{-6}$. InGaAs/InP APDs are cooled down to -40$^\circ$C, using Peltier modules. Single photon detectors were realized with both sinusoidal wave gating and a self-differencing circuit.

Fast key distillation is realized with standard PCs only in software implementation, using the improved algorithm for privacy amplification. For error correction, the low density parity check (LDPC) code is adopted to achieve a performance approaching Shannon's limit. For privacy amplification, the calculation time is reduced from $O(n^2)$ to $O(n\log(n))$ for the block size $n$ by using the fast Fourier transform algorithm for multiplying the Toeplitz matrix and a reconciled key. The reduction amounts to 4 orders of magnitudes for $n=10^6$, which is currently known to be the minimum block size to eliminate the finite size effect in distilling the secure key. The secure key rate was 2 kbps and the QBER is about 4.5%. We confirmed the stability of the key generation. Figure 12 shows the experimental results of continuous operation. Stable key generation for about 3 days was demonstrated.

We also developed an OTP smartphone using QKD, whose photo and diagram are shown in Fig. 13. It provides end-to-end encryption of voice data between smartphones. A secure key is downloaded from the accompanying QKD devices to the smartphone. Voice data is encoded with a rate of 1 kB per second, which requires approximately 1.2 MB for a 10 min bidirectional talk. With a 2 GB Secure Digital (SD) card, continuous conversation for 10 days by OTP encryption can be supported with a single downloading. Once a key has been used for encryption, it is immediately eliminated from any memory inside the smartphone.

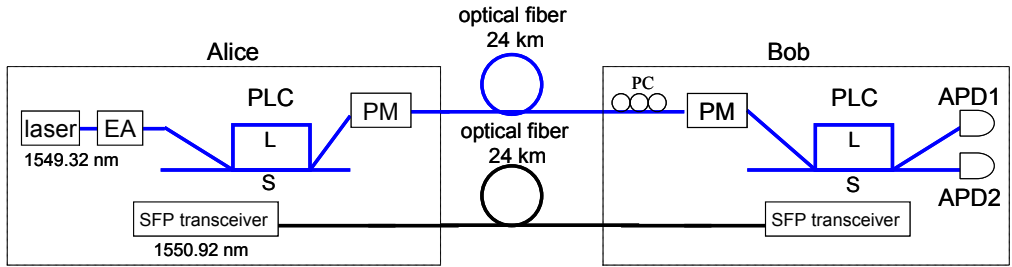

Fig.11. Schematic diagram of the QKD setup. EA: electro-absorption modulator; PLC: planar light circuit (L: long arm; S: short arm); PM: phase modulator; PC: polarization controller; APD: InGaAs/InP avalanche photodiode; SFP transceiver: small form factor pluggable transceiver.

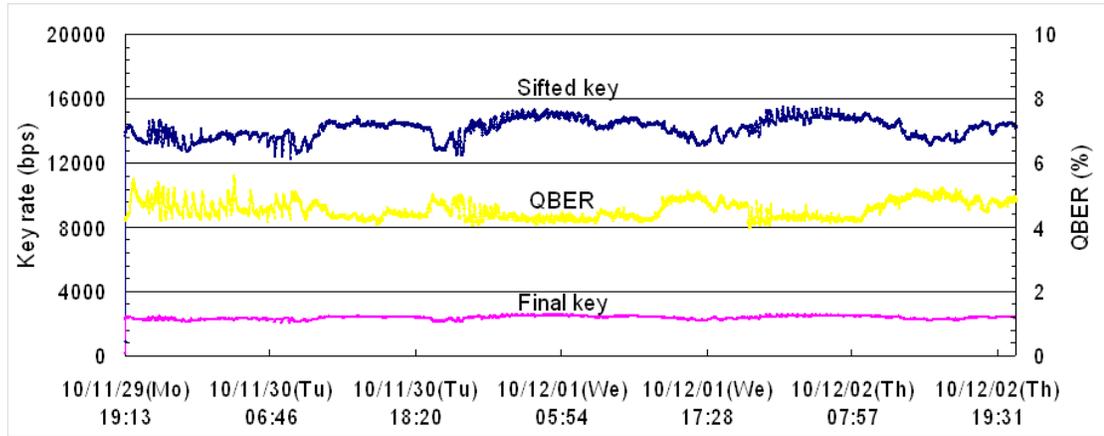

Fig. 12. Measured quantum bit error rate (QBER), and sifted and final key rates over a duration of about three days.

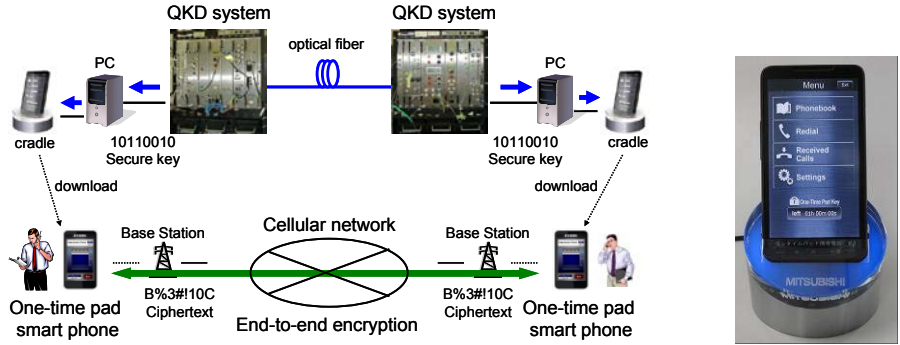

Fig. 13 Schematic diagram of the QKD system with one-time pad smartphones.

## 3.5. IDQ system

IDQ demonstrated stable long-term operation in the 13 km link using their commercial product called Cerberis. A photograph of the system that was installed at the Otemachi node is shown in Fig. 14. The system is working in a phase coding configuration and is based on the so-called Plug & Play optical platform. This is a go and return configuration which allows high quality auto-compensation of polarization and phase fluctuations of the quantum channel.

This device uses SARG04 protocol. SARG04 protocol differs from the standard BB84 protocol by the sifting part only, but this small difference allows SARG04 protocol to be more robust against photon number splitting attacks.

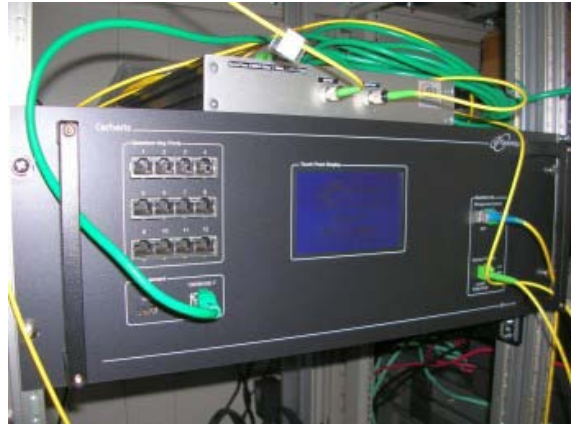

Fig. 14. Photograph of the Cerberis at Otemachi with the optical filter on top.

The system has been installed in only a few hours in March. It has been running continuously for half a year except for a two months period. During this period of time, either the quantum channel was not available or IDQ was testing its system with an additional filter. This filter had to be added to reduce the additional noise due to cross-talk effects between the quantum and classical channels in neighboring optical fibers. Figure 15 shows a typical data of secure key rate as a function of time over a 45 days period. In this figure, data is shown from 27$^{th}$ June 2010 till the day of the demonstration in the UQCC2010 on 18$^{th}$ October 2010 [34]. The working of the system between March and the end of June is not represented because this period of time was used to tune the system in order to optimize the secure key rate. This optimization consists in changing the single photon detector parameters in order to find a good trade-off between high efficiency values and low afterpulsing probabilities. In Fig. 15, we can see that the secure key rate was about 300-350 bps before August. After inserting the additional filter, the secure key rate increased to 400-420 bps. This shows that whereas the quantum channel link loss increased due to the addition of the filter, the reduction of the additional noise allows the QKD system to exchange more secure keys.

In terms of QBER, the value goes from 4% down to approximately 2% after the addition of the filter. Crosstalk noise can impose a strong limitation on the performance of QKD if it is not mitigated by spectral filtering. The sifted key rate was about 1.5kbps and PA was done by Toeplitz matrices with an input bit-string size of about 500 kbit.

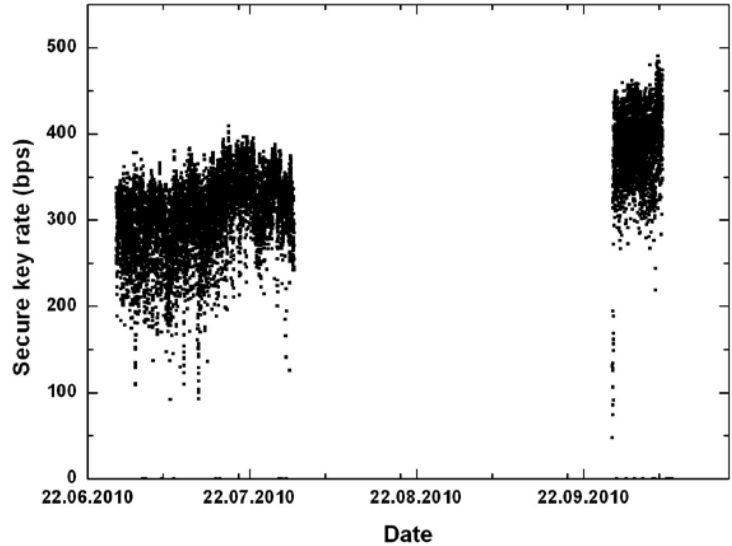

Fig. 15. Secure key rate recorded for a few month period of the IDQ system.

*3.6 All-Vienna system*

The All-Vienna team operated the entanglement based QKD employing the BBM92 protocol [19] between nodes Koganei-2 and Koganei-3. In this scheme the state is not actively prepared by modulators (as in the case of BB84, decoy states QKD or other "prepare-and-measure" schemes), but rather an entangled state is measured by passive polarization analyzers situated in the spatially separated devices of Alice and Bob as shown in Fig. 16 [see also Ref. 35]. Thereby quantum correlations are transferred into secrets. A significant advantage of this setup over non-entanglement schemes is the robustness against certain side-channel attacks. A particular example is the case of an accidental increase in fluctuation of the laser power causing a change in the photon statistics, which typically remains unknown to Alice and Bob in a prepare-and-measure scheme. In the presented setup an increased pair generation rate causes more double clicks at Alice, which after appropriate "squashing" procedures are translated into an increased QBER and key rate reduction, but does not lead to security leaks. Further, a direct control of the detectors by the adversary is disallowed by monitoring the incident power down to microwatts, which prevents blinding the detectors remotely. Moreover, to prevent leakage of timing information, individual delay and timing modules have been installed at Alice and Bob.

The entanglement source at Alice produces two photons of different wavelengths (810 nm and 1550 nm) by spontaneous parametric down-conversion in a ppKTP crystal set. The 810 nm photon is measured by Alice in a passive polarization analyzer (one 50/50 beam splitter and two polarizing beam splitters) and detected by four Si-APDs (avalanche photo diodes). The 1550 nm photon of the pair travels down the quantum channel with low transmission losses and is registered by Bob using a fiber based passive polarization analyzer with four InGaAs-APDs for detection. Trigger pulses (1610 nm) are generated at Alice and multiplexed on the quantum channel to gate the InGaAs-APDs at Bob.

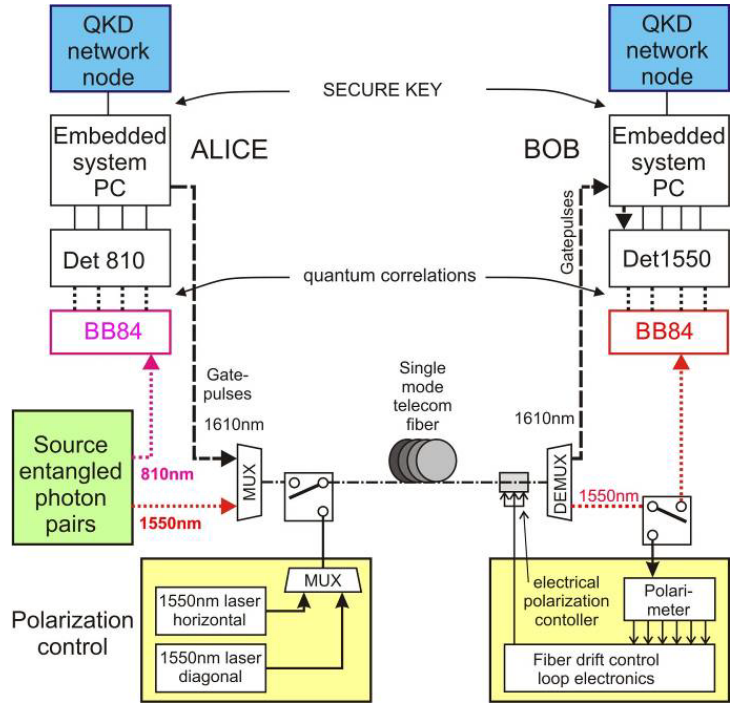

Fig. 16. Block scheme of a polarization entanglement based QKD setup.

The measurement results at Alice and Bob are further processed by an FPGA and an embedded PC (per device), delivering secure key over predefined interfaces. Additional FPGA electronics controls the source stabilization module at Alice and, together with a polarimeter, the polarization control module at Bob. To compensate the unavoidable polarization drift of fibers, 1550 nm classical light pulses in horizontal and diagonal polarization are launched through the optical path routinely every 10 minutes or on demand to counteract high QBER. Polarization drifts along the fiber are detected and compensated at the receiver by a sophisticated polarization control algorithm. The system is also equipped with a relatively slow but very efficient state alignment mechanism which ensures robustness against strong temperature changes of the environment of Alice and Bob or temporal loss of fiber connectivity. The resulting errors are detected and normal functionality is restored. The system was miniaturized to fit in a 19 inch square box, including detectors, electronics and feedback systems to control the operation.

The software of the All Vienna Prototype deployed in the Tokyo QKD Network has been upgraded significantly since the version used in the SECOQC Entangled System [35]. Further modularization allows the QKD system to seamlessly operate within either a SECOQC-style network or the Tokyo QKD Network setup without changes to the underlying QKD process. The source of random numbers used by the system is now configurable and uses by default a quantum number generator from IQOQI.

Specifically, QKD post-processing involves the standard stages of sifting, reconciliation (error correction), confirmation and privacy amplification. Due to the relatively law key generation rate (see below) the application of both LDPC and CASCADE error correction strategies was feasible. We have chosen to employ the CASCADE-reconciliation module in the parallel CASCADE flavor, proposed by L. Salvail during SECOQC (i.e. communication between Alice and Bob is not performed bit-wise but protocol messages are grouped and maximally postponed), allowing for significant reduction of communication latency and real-time error correction speed. The block length of the privacy amplification module, essential as a consequence of the "finite size effects", discussed in previous sections, is configurable and was set to 300 kbit. Privacy amplification is based on a 2-universal hash function family realized as binary matrix multiplication with Toeplitz matrices. Direct application is computationally ineffective but this problem is easily alleviated by observing that Toeplitz matrix multiplication is a convolution and can therefore be speeded up by Fourier transform (see discussion in Section 3.4 above). In view of the fact that the Fourier transform of binary functions is needed, Number Theoretic Transform is used over a prime p-field, with $p=13*2^{20}+1$.

A major goal of the software development process was to evolve the code used in the previous generation of the system into a generic framework that can be easily adapted to other QKD implementations. The AIT team has released this software with an open-source license to other independent groups to be validated, tested and further developed.

Figure 17 represents a typical overnight measurement to demonstrate the stability of the system with artificially increased loss over additionally available deployed fiber loops and a short fiber spool in the lab of the Tokyo prototype. We have deliberately chosen to show here a partly misaligned state of the source in the early stages of the operation. In result we get an increase of the relative share of dark counts and high error rate of 5% to 7%. Correspondingly the average sifted bit rate is 0.93 kbps and the secure key rate is low being around 0.25 kbps as compared to the SECOQC performance of the system producing secure key at a rate of 2 to 3 kbps [35]. However the graph serves well to illustrate the stability of the system. Temperature change of the environment can compromise polarization stability of the arms of Bob's BB84 module leading to a slow decrease of the secure key rate.

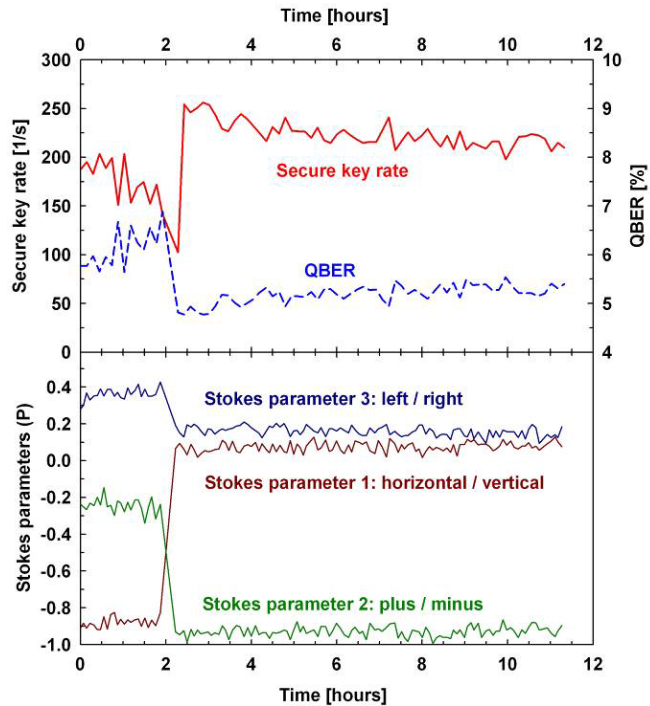

Fig. 17. Key rate, QBER and polarization stability during 12 hours of operation in the Tokyo prototype.

The QKD-system autonomously counteracts these problems (see Fig. 17 at hour 2) by an automatic launch of the state alignment mechanism, which uses the quantum correlations to optimize the alignment. The QBER is reduced and the secure key rate is optimized (upper graph in Fig. 17). As a side effect, a significant polarization rotation registered by the polarimeter used for the routine polarization control mechanism can be observed (lower part of Fig. 17). During the next 8 hours the Stokes parameters remain constant indicating the full control of the polarization drift of the fiber by the periodical polarization control cycles. Still, the clearly visible long term drift of the QBER is monitored to repeat, if necessary, state alignment procedures in order to maintain a stable operation over weeks and months.

## 4. Demonstration of secure network operation

The live demonstration of secure TV conferencing, eavesdropping detection, and rerouting of QKD links on the Tokyo QKD Network was performed and made public in October 2010 [36]. The configuration of secure TV conferencing is depicted in Fig. 18. In the secure TV conferencing, Polycom Video Conference Systems were set up in Koganei-1 and Otemachi-2 which were directly connected via the JGN2plus L2-VPN. A live video stream for this VPN was encrypted by OTP at the KMAs in the stored key mode. The encryption rate was 128 kbps. Secure keys were provided by either of two QKD relay routes. One is via Koganei-2 with a total distance of 135 km, and the other is via Otemachi-1 with a total distance of 69 km, as shown by the red and blue lines, respectively. The former one was used as the primary route. Figure 19(a) shows a screenshot of the KMS screen, indicating that the 90 km QKD link works flawlessly. The link was then attacked by intercepting a photon stream from the fiber via a high reflective mirror and injecting a CW laser with the same power as the tapped one into the fiber. The KMS detected this attack in a few seconds as the result of a sudden increase of the QBER, stopped the QKD process in the link, and raised an alarm. Figure 19(b) is a copy of the KMS screen shortly after the detection of the attack. The KMAs at Koganei-1 and Koganei-2 had some stored secure key material, and the secure TV conferencing could continue for a while. The KMS immediately switched to the secondary route to be able to continue with the key distribution before the key buffer ran out of key. Thus secure TV conferencing could go on unimpeded and security was guaranteed.

A key relay was also tested and operated successfully not only for the above secure video streaming but also in testing various relay routes including nodes Koganei-3 and Hongo with the teams of All Vienna and IDQ.

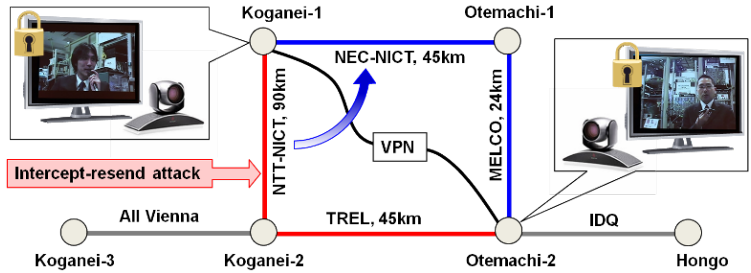

Fig. 18. The network configuration for secure TV conferencing between Koganei and Otemachi. The VPN used for video transmission can choose between two relaying QKD routes, i.e. the red line via Koganei-2 and the blue line via Otemachi-1.

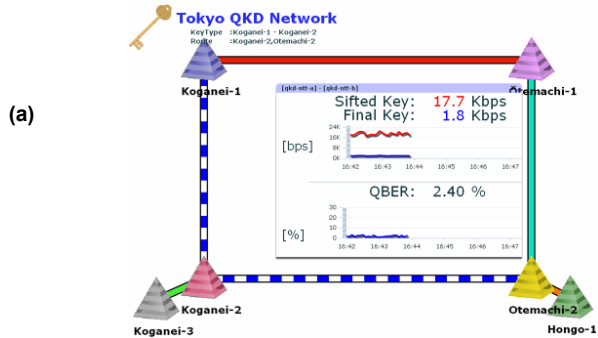

(a)

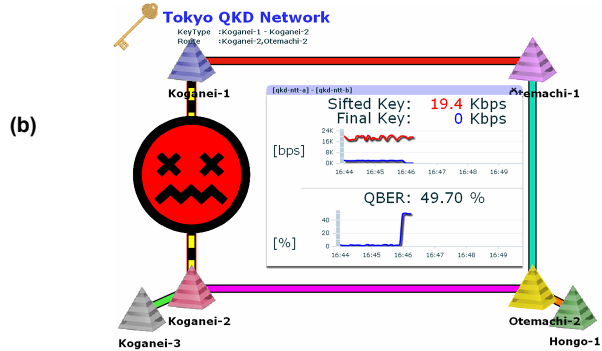

Fig. 19. Screenshots of the KMS
(a) before the attack, and (b) right after the KMS detected the attack.

## 5. Conclusions and Future Outlook

We have presented the results of a QKD field trial using the Tokyo QKD Network. In this trial, novel GHz-clocked QKD systems, a QKD smartphone, a reliable commercial QKD product, and an entanglement QKD system were employed. They were interconnected via several key management agents using a common API, and managed by a single key management server. The demonstrated applications include secure TV conferencing using QKD-OTP and QKD smartphone. The former application was supported by two relayed QKD routes, each of which includes one trusted relay node, and by a rerouting function to switch from a hacked to an alternative secure link when an eavesdropper in the system was detected. These demonstrations suggest that practical applications of QKD in a metropolitan network may be just around the corner.

The near-term use case is likely to be high-end security applications that have been relying up to now on trusted couriers for key exchange. To become a practical solution long-term reliability of QKD needs to be guaranteed. This includes not only stable operation but also security assurance over a long span of time.

Many QKD protocols, such as the one-way BB84 protocol, have been proven to be unconditionally secure, which means the *protocol* cannot be 'cracked' as long as the laws of physics remain true. On the other hand real world implementations have unavoidable imperfections and will therefore be susceptible to side-channel attacks. To guarantee the security of real world implementations it is important to characterize and define the underlying assumptions. Furthermore, possible side channels should be investigated and appropriate countermeasures devised. In this regard, test runs of the Tokyo QKD Network need to be continued to discover and characterize the implementation loopholes and incorporate appropriate countermeasures. Sharing information on such activities with various QKD network testbeds around the world will establish a common denominator for standardizing QKD in the context of security certification. A strong QKD network should incorporate active stabilization schemes and as many side-channel countermeasures as possible without sacrificing performance.

We would like to further address three future issues: The first one is to develop a more efficient key management technology to allow a scalable extension to multipoint network configurations with multicasting applications. Introducing a new node processing method based on network coding in the communication layer may also be helpful to reduce the consumption of secure key.

The second one is to integrate QKD network technologies into an emerging infrastructure of optical communications, called photonic network, where optical signals are processed within the optical domain without conversion to electronic form. Quantum networking should eventually be made in the WDM infrastructure of optical path networks, which corresponds to the very bottom layer of photonic networks. According to user requests, an efficient QKD link will be realized directly via optical cross connects and reconfigurable optical add-drop multiplexers. As the range of quantum transparency gets wider, secure keys can be used to tightly encrypt the control signals in the control plane. Actually, novel control technologies such as the generalized multi-protocol label switching make the control plane more open to the upper layers, other operators and end users. Malicious hackers may also have the opportunity to easily access the control plane which could seriously endanger the security of the entire network. QKD will play a key role in developing secure photonic networks.

The last one is to widen QKD applications not only to protect data confidentiality but also to provide services, which are essential functions in current security systems such as message authentication, identification and digital signature. While QKD readily supports "unconditionally secure" message authentication, identification and digital signature require further research and may be realized using existing QKD hardware at the expense of assuming an adversary bounded by finite quantum memory or additional (next generation) quantum resources available to the legitimate parties such as quantum processing and quantum memory. In any case, if QKD performance is further improved and cost reduced, then prospective QKD networks (featuring the functionalities discussed above) could become an essential infrastructure for secure key generation for a wide variety of cryptographic objectives. This may be the major impetus to pursue the improvement of QKD technology and future QKD network research.

**Acknowledgements**

This work was supported on the Japanese side by the project on "Research and development for practical realization of quantum cryptography" of NICT under the Ministry of Internal Affairs and Communications of Japan. IDQ and TREL would like to acknowledge partial financial support through the EU FP7 projects Q-ESSENCE and QuReP. Also, the European project SECOQC is acknowledged for contributing to the development of equipment used in the Tokyo QKD network.

Finally, the authors express their thanks to Kazuhiko Nakamura, Kenji Terada, Nobuo Kawashima, Akira Hanzawa, Ryota Saito, Ayako Kikuchi and Hiroshi Hayano for their support in setting up the Tokyo QKD Network. The authors from the All Vienna team acknowledge the significant contribution of Thomas Themel and Andreas Happe for software development under the supervision of Oliver Maurhart, and express sincere thanks to Michael Hentschel, Christoph Pacher, Thomas Lorünser and Rupert Ursin for continuous development and logistic support as well as for helpful discussions throughout the project.